\documentclass[a4paper,fleqn,usenatbib]{mnras}

\usepackage{newtxtext,newtxmath}

\usepackage[T1]{fontenc}
\usepackage{ae,aecompl}

\usepackage{graphicx}	
\usepackage{amsmath}	
\usepackage{amssymb}	
\usepackage{multirow}
\usepackage{array}
\usepackage{subfig}

\title[Galaxy Classification with CNNs]{Galaxy Morphology Classification with Deep Convolutional Neural Networks}

\author[J. M. Dai et al.]{
Jia-Ming Dai,$^{1,2}$ \thanks{E-mail: daijiamingdl@gmail.com}
Jizhou Tong$^{1}$ 
\\
$^{1}$ National Space Science Center, Chinese Academy of Sciences, Beijing 100190,China\\
$^{2}$ University of Chinese Academy of Sciences, Beijing 100049, China\\
}

\date{Accepted XXX. Received YYY; in original form ZZZ}

\pubyear{2018}

\begin{document}
\label{firstpage}
\pagerange{\pageref{firstpage}--\pageref{lastpage}}
\maketitle

\begin{abstract}
We propose a variant of residual networks (ResNets) for galaxy morphology classification. The variant, together with other popular convolutional neural networks (CNNs), are applied to a sample of 28790 galaxy images from Galaxy Zoo 2 dataset, to classify galaxies into five classes, i.e. completely round smooth, in-between smooth (between completely round and cigar-shaped), cigar-shaped smooth, edge-on and spiral. A variety of metrics, such as accuracy, precision, recall, F1 value and AUC, show that the proposed network achieves the state-of-the-art classification performance among the networks, namely, Dieleman, AlexNet, VGG, Inception and ResNets. The overall classification accuracy of our network on the testing set is 95.2083\% and the accuracy of each type is given as: completely round, 96.6785\%; in-between, 94.4238\%; cigar-shaped, 58.6207\%; edge-on, 94.3590\% and spiral, 97.6953\% respectively. Our model algorithm can be applied to large-scale galaxy classification in forthcoming surveys such as the Large Synoptic Survey Telescope (LSST).
\end{abstract}




\section{Introduction}\label{sec:intro}

Galaxies have various shapes, sizes and colors. To understand how these morphologies of galaxies relate to the physics that create them, galaxies need to be classified. Thus galaxy morphology classification is a key step to the study on galaxy formation and evolution. In 1926, Edwin Hubble first proposed the \lq\lq Hubble Sequence\rq\rq~  using visual inspection with fewer than 400 galaxy images (also called \lq\lq Hubble Tuning Fork\rq\rq), classifying galaxies into three basic types: elliptical, spirals and irregular \citep{hubble1926extragalactic,sandage2005classification}. And \lq\lq Hubble Sequence\rq\rq~ is still in use today. For a long time, astronomers used the visual inspection to classify galaxies and update Hubble' classification scheme. In recent decades, large scale surveys such as the Sloan Digital Sky Survey (SDSS) have resulted a huge amount of galaxy images. Classifying these huge images by astronomers is not only time consuming but also a impossible mission.

Then Galaxy Zoo project attempted to solve the problem and was launched \citep{lintott2008galaxy,lintott2010galaxy}. Galaxy Zoo 1 with a dataset made of a million galaxy images by the Sloan Digital Sky Survey, invited a large number of citizen scientists to provide the basic morphological information and identify if a galaxy was \lq\lq spiral\rq\rq, \lq\lq elliptical\rq\rq, \lq\lq a merger\rq\rq or \lq\lq star/don't know\rq\rq \citep{lintott2008galaxy}. The project achieved a huge success that the million galaxy images were annotated within several months. And then Galaxy Zoo 2 \citep{willett2013galaxy}, Galaxy Zoo: Hubble \citep{willett2016galaxy}, and Galaxy Zoo: CANDELS \citep{simmons2016galaxy} are launched respectively. Unfortunately, this approach still doesn't keep up with the pace of data growth. Astronomers turn their sights to a automatic classification method.

Galaxy morphology classification using machine learning methods has played an important role in the past 20 years. Artificial neural networks, Naive Bayers, decision tree and Locally weighted Regression have been applied in galaxy classification on relatively small datasets in early work \citep{naim1995automated,owens1996using,bazell2001ensembles,de2004machine}. \citet{de2004machine} found that the accuracy dropped from 95.66\% to 56.33\% classifying galaxies into 2 classes to 5 classes. \citet{banerji2010galaxy} used artificial neural networks to assign galaxies to 3 classes with several input parameters, e.g., colors, shapes, concentration and texture. \citet{gauci2010machine} used decision tree and fuzzy logic algorithms to galaxy morphology classification based on the designed photometric parameters and spectra parameters. \citet{ferrari2015morfometryka} measured galaxy morphological parameters including Concentration, Asymmetry, Smoothness, Gini coefficient, Moment, Entropy and Spirality to automatically classify galaxies employed the Linear Discriminant Analysis (LDA).  Other recent galaxy classification methods \citep{orlov2008wnd,huertas2011revisiting,polsterer2012galaxy} all need feature extraction, which needs human careful design. It is well known that the performance of classification depends on the choice of data representation, called feature engineering \citep{lecun2015deep}. Feature engineering needs domain expertise and is time-consuming.

In the past three years galaxy morphology classification using deep learning algorithms has obtained more attention. Deep learning models are composed of multiple non-linear layers to learn data representation, which allow to be fed with raw data directly and automatically learn the representations of data \citep{bengio2013representation,lecun2015deep}.  After multiple non-linear transforming, the representations of higher layers are abstract and beneficial for discrimination and classification. Deep convolutional neural networks (CNNs) have become the dominant approach in image classification task. With the availability of the large number of Galaxy Zoo labeled dataset, some works have yielded good results. \citet{dieleman2015rotation} for the first time used a 7-layers CNN to galaxy morphology classification which exploits galaxy images translation and rotation invariance. Then, \citet{gravet2015catalog} used the \citet{dieleman2015rotation} model to classify high redshift galaxies in the 5 Cosmic Assembly Near-infrared Deep Extragalactic Legacy Survey (CANDELS).  \citet{hoyle2016measuring} used CNNs to estimate the photometric redshift of galaxies.  \citet{kim2016star} presented a star-galaxy classification framework similar to VGG  \citep{simonyan2014very}. Recently, CNNs have been applied to find strong gravitational lenses in the Kilo Degree Survey \citep{petrillo2017finding}.  And \citet{aniyan2017classifying} used CNNs to classify radio galaxies into FRI, FRII and Bent-tailed radio galaxies.

In this study, we propose a modified residual network (ResNet) for galaxy morphology classification. We select 28790 galaxy images from Galaxy Zoo 2 dataset and use five forms of data augmentation to enlarge the number of our training samples in data preprocessing to avoid overfitting. The variant we proposed combines the advantages of Dieleman model   \citep{dieleman2015rotation} and residual networks. In addition, We implement several other popular CNNs models, including Dieleman, AlexNet \citep{krizhevsky2012imagenet}, VGG \citep{simonyan2014very}, Inception \citep{szegedy2015going, ioffe2015batch, szegedy2016rethinking, szegedy2017inception} and ResNets \citep{he2016deep, he2016identity} and systematically compare the classification performance of ours with these CNNs model. As expected, we demonstrate that our model achieves a state-of-the-art performance. Furthermore, to understand what the CNNs learn, we visualize the filters weights and feature maps to give a qualitative empirical analysis.

This paper is organized as follows. We introduce the dataset selection in Section \ref{sec:dataset}. Section \ref{sec:cnns} describes deep learning and convolutional neural networks (CNNs).  Section \ref{sec:approach} contains data preprocessing pipeline, data augmentation, the residual network we have proposed and the training tips. Section \ref{sec:results}
are the results and analysis of our network and other CNNs models. Finally, we draw conclusions and future work in Section \ref{sec:conclusions}.

\section{Dataset} \label{sec:dataset}

The galaxy images in this study are drawn from Galaxy Zoo-the Galaxy Challenge \footnote{https://www.kaggle.com/c/galaxy-zoo-the-galaxy-challenge}, which contain 61578 JPG color galaxy images with probabilities that each galaxy is classified into different morphologies. Each image is of $424\times424\times3 $ pixels in size taken from the Galaxy Zoo 2 main spectroscopic sample from SDSS DR7 \footnote{http://www.sdss.org/}. The morphological classifications vote fractions are modified version of the weighted vote fractions in the Galaxy Zoo 2 project \footnote{https://www.galaxyzoo.org/}. The classifications vote fractions have high level of agreement and authority with professional astronomers \citep{willett2013galaxy}. The data has been used in studies of galaxy formation and evolution \citep{land2008galaxy,schawinski2009galaxy,bamford2009galaxy,willett2015galaxy}.

\begin{table*}
	\centering
	\caption{Clean samples selection in Galaxy Zoo 2. The clean galaxy images are selected from Galaxy Zoo 2 data release \citep{willett2013galaxy}, in which thresholds determine well-sampled galaxies. And here they are called clean samples. Thresholds depend on the number of votes for a classification task considered to be sufficient. As an example, to select the spiral, cuts are the combination of $f_{features/disk} \geq 0.430$ , $f_{edge-on,no} \geq 0.715$, $f_{spiral,yes} \geq 0.619$.}
	\label{tab:dataselection}
	\begin{tabular}{clcll}
		\hline\hline
		Class & Clean sample & Tasks & Selection & $N_{sample}$ \\
		\hline
		0 & Completely round smooth & T01  & $f_{smooth} \geq 0.469$   & 8434 \\
        & & T07 &$f_{completely ~round} \geq 0.50$ & \\
        \hline
        1 & In-between smooth & T01  & $f_{smooth} \geq 0.469$   & 8069 \\
        &  &T07 &$f_{in-between} \geq 0.50$ & \\
        \hline
        2 & Cigar-shaped smooth & T01  & $f_{smooth} \geq 0.469$   & 578 \\
        &  & T07&$f_{cigar-shaped} \geq 0.50$ & \\
        \hline
        3 & Edge-on  & T01  & $f_{features/disk} \geq 0.430$   & 3903 \\
        & &T02 & $f_{edge-on,yes} \geq 0.602$ & \\
        \hline
        &   & T01  & $f_{features/disk} \geq 0.430$    &  \\
        4& Spiral &T02 &$f_{edge-on,no} \geq 0.715$ & 7806 \\
        & & T04 &$f_{spiral,yes} \geq 0.619$ & \\
        \hline

	\end{tabular}
\end{table*}

\begin{figure}
  \centering
  \includegraphics[scale=0.7]{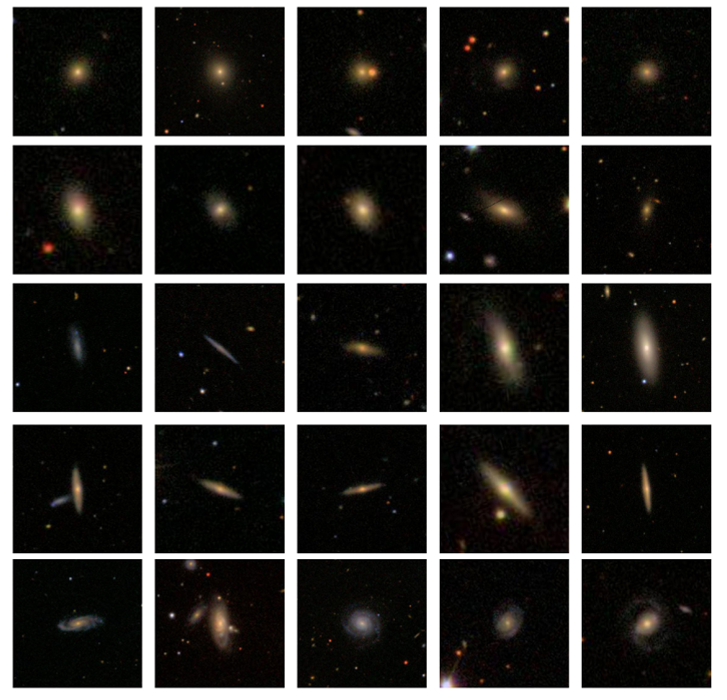}\\
  \caption{Example galaxy images from the dataset. Each row represents a class. From top to bottom, their Galaxy Zoo 2 labels are: completely round smooth,
    in-between smooth, cigar-shaped smooth, edge-on and spiral. They are referred to as 0, 1, 2, 3 and 4.}
  \label{fig:galaxy}
\end{figure}

In this study clean samples are selected that match a specific morphology category with their appropriate thresholds \citep{willett2013galaxy}, which depend on the number of votes for a classification task considered to be sufficient. For example, to select the spiral, cuts are the combination of $f_{features/disk} \geq 0.430$ , $f_{edge-on,no} \geq 0.715$, $f_{spiral,yes} \geq 0.619$. These thresholds are considered conservative to select clean samples in \citet{willett2013galaxy} . By this means, we assign galaxy images to five classes, i.e. completely round smooth, in-between smooth(between completely round and cigar-shaped), cigar-shaped smooth, edge-on and spiral. In practice, all thresholds are derived from \citet{willett2013galaxy} except thresholds of smooth galaxy are loosened from 0.8 to 0.5, and full details refer to \citet{willett2013galaxy}. Table \ref{tab:dataselection} shows the clean samples selection criterion for every class. The 5 classes galaxies are referred to as 0, 1, 2, 3 and 4, each contains a sample of 8434, 8069, 578, 3903 and 7806 respectively. Figure \ref{fig:galaxy} shows the galaxy images randomly selected from the dataset and each row represents a class. From top to bottom, their labels are: 0, 1, 2, 3 and 4 respectively.

\begin{figure}
\centering
\includegraphics[width=0.5\textwidth]{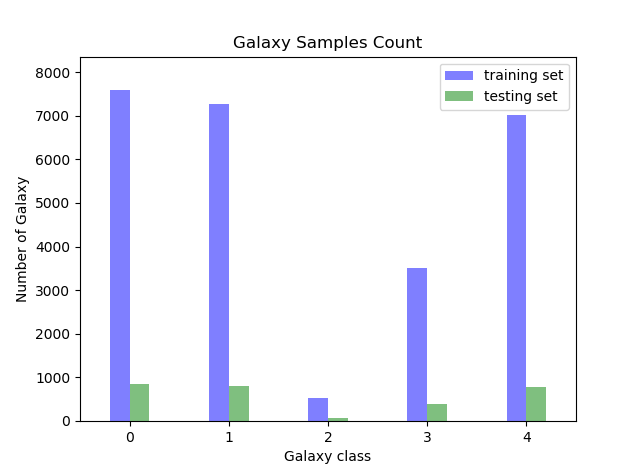}\\
\caption{Galaxy Samples Counts}\label {fig:bar}
\end{figure}

The dataset reduces to 28790 images after filtering, then is divided into training set and testing set by a ratio of 9:1. Thus there are 25911 images for training set to train our model and remaining 2879 images for testing set to evaluate our model. Training set and testing set have the same distribution. Table \ref{tab:samples} gives the number of galaxy images in each morphological class of training set and testing set and Figure \ref{fig:bar} reproduces the dataset graphically.

\begin{table}
\centering
\caption{Number of galaxy images in each morphological class in each set. 0, 1, 2, 3, 4 represent completely round, in-between, cigar-shaped, edge-on and Spiral, respectively.}
\label{tab:samples}
\begin{tabular}{ccccccc}
\hline\hline
& 0& 1&2&3&4&Total \\
\hline
Training set & 7591 & 7262 & 520 & 3513 &7025 & 25911 \\
Testing set & 843 & 807 & 58 & 390 &781 & 2879 \\
\hline
Data set & 8434 & 8069 & 578 & 3903 &7806 & 28790 \\
\hline
\end{tabular}
\end{table}

\section{Deep Convolutional Neural Networks} \label{sec:cnns}

Deep learning models are composed of multiple layers to automatically learn data representations from the raw data, which are capital for classification, localisation, detection, segmentation without feature extraction \citep{lecun2015deep}. Deep convolutional neural networks (CNNs) have played an important role in deep learning \citep{goodfellow2016deep}. Convolutional neural networks have become the dominant approach in image classification. In this section, we briefly introduce artificial neural networks (ANN), convolutional neural networks (CNNs), especially residual networks (ResNets).

\subsection{Artificial Neural Networks}\label{subsec:ann}

\begin{figure}
  \centering
  \includegraphics[scale=0.7]{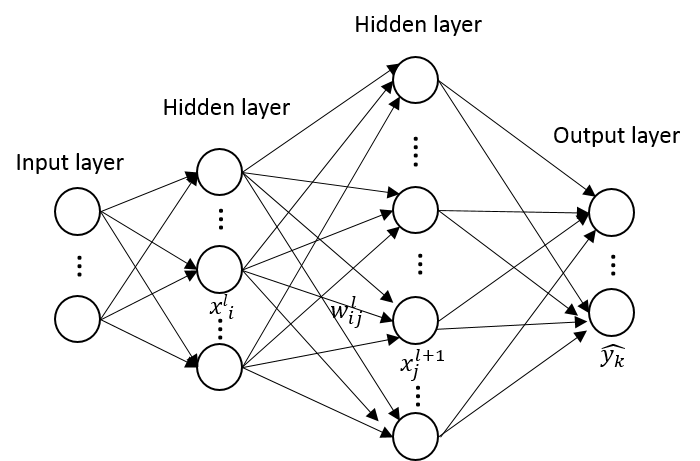}\\
  \caption{Schematic of a feed forward neural network.}\label{fig:ann}
\end{figure}

Artificial neural networks (ANN) are made up of simple adaptive units  interconnected, which can simulate biological nervous system, interaction in response to real world objects  \citep{KOHONEN19883}. Figure \ref{fig:ann}  shows a simple feed forward neural network. It is composed of input layer, hidden layer and output layer. Formally, define $ x_{i}^{l},~x_{j}^{l+1}$ as the $i$-th neuron of $ l$-th layer, the $j$-th neuron of $ (l+1)$-th layer, define $ w_{ij}^l, b_j^l $ as weights, bias of the $ l$-th layer, respectively. Then, the outputs of the $ l$-th layer are $  x_{j}^{l+1} $:

\begin{equation}\label{eq:wb}
  x_{j}^{l+1}=f(\sum_{i\in N^{l}}(w_{ij}^{l}x_{i}^{l}+b_{j}^{l}))
\end{equation}

where $N^{l}$ is the number of $l$-th layer, $ f $ is the activation function. Activation functions have many types, such as the popular rectified linear unit (ReLU)  \citep{nair2010rectified}, $ f=\max(0,x)$ , sigmoid, tanh, Leaky ReLU, ELU and so on.

Then, let  $ \hat{y}=( \hat{y_1},\hat{y_2} , \cdots, \hat{y_k}, \cdots, \hat{y_m})$ be the output of network, $ y=( y_1, y_2, \cdots, y_k, \cdots y_m ) $ be the desired output and we can define a cost function $\ell( \hat{y},y)$. In classification task, the cost function can be a cross entropy. Especially, in binary classification, the cross entropy can be defined as

\begin{equation}\label{eq:loss}
  \ell(\hat{y},y)=-y\log\hat{y} -(1-y)\log(1-\hat{y}).
\end{equation}
where $ y\in \{0, 1\}, ~ \hat{y}\in[0, 1] $. Then, we need to compute cross entropy of all training data. In order to minimum the cross entropy, we use stochastic gradient descent (SGD) to update the weights and bias until the loss function converge:

\begin{equation}\label{sgdw}
  w_{n+1}^l=w_n^l-\eta\frac{\partial\ell}{\partial w_n^l}.
\end{equation}
\begin{equation}\label{sgdb}
  b_{n+1}^l=b_n^l-\eta\frac{\partial\ell}{\partial b_n^l}.
\end{equation}
Where $ \eta $ is learning rate. Of course, now we generally use mini-batch stochastic gradient descent instead of all data stochastic gradient descent in practice in order to save training time, which can seek for a local optimal solution.

\subsection{Convolutional Neural Networks}\label{subsec:cnn}

Convolutional Neural Networks (called CNNs or Convnets) \citep{Cun1989Handwritten} are designed to process multiple arrays data, for example, image data. CNNs have become very successful in practical applications. A classical layer of CNN is made up of three stages. In first stage, the layer performs several convolutions. Then, a non-linear activation function such as ReLU is applied. At last, a pooling function modifies the output the layer \citep{goodfellow2016deep}. CNNs generally contain convolutional layers, pooling layers and fully connected layers.

\textbf{Convolutional layers}. Convolution is a specialized kind of linear operation. Discrete convolution can be viewed as multiplication by a matrix. Convolutional layers can be computed by

\begin{equation}\label{eq:conv}
  x_j^l=f(\Sigma_{i\in M_j}x_i^{l-1}\times k_{ij}^l+b_j^l).
\end{equation}
Where $l$ is the number of layer, $f$ is activation function usually ReLU,  $k$ represents convolutional kernel, $M_j$ represents the receptive field and $b$ is bias.

\textbf{Pooling layers} (also called \textbf{subsampling}). Pooling can be achieved by taking average (average pooling) or taking maximum ( max pooling) within a rectangular neighborhood. For an image, it can reduce the size of images.

\textbf{Fully connected layers}. Fully connected layers are usually followed by the last pooling layer or the convolutional layer, and every neuron in fully connected layers is connected to all the neurons in the upper layers.

Generally, convolutional networks (CNNs) have sparse connectivity, parameter sharing and equivariant representation, three important ideas.

Deep convolutional networks (CNNs) have brought a series of breakthroughs in image classification. And CNNs are getting deeper and deeper, from 8 layers \citep{krizhevsky2012imagenet}, 16/19 layers \citep{simonyan2014very}, 42 layers \citep{szegedy2016rethinking}, to 152 layers \citep{he2016deep}. In order to train deeper networks, some new techniques are adopted, such as ReLU \citep{nair2010rectified}, dropout \citep{srivastava2014dropout}, GPUs, data augmentation \citep{krizhevsky2012imagenet}, batch normalization (BN) \citep{ioffe2015batch} and so on. Now CNNs models have developed several versions, primarily including AlexNet \citep{krizhevsky2012imagenet}, VGG \citep{simonyan2014very}, Inception \citep{szegedy2015going, ioffe2015batch, szegedy2016rethinking, szegedy2017inception}, ResNets \citep{he2016deep,he2016identity} and DenseNet \citep{huang2016densely}.

\subsection{Residual Networks}\label{subsec:resnet}

Deep residual networks (ResNets) are reported in \citet{he2016deep,he2016identity}, which can deepen the networks up to thousands of layers and achieve state-of-the-art performance. In this section, we give a brief description of ResNets.

\begin{figure}
\centering
\includegraphics[scale=0.5]{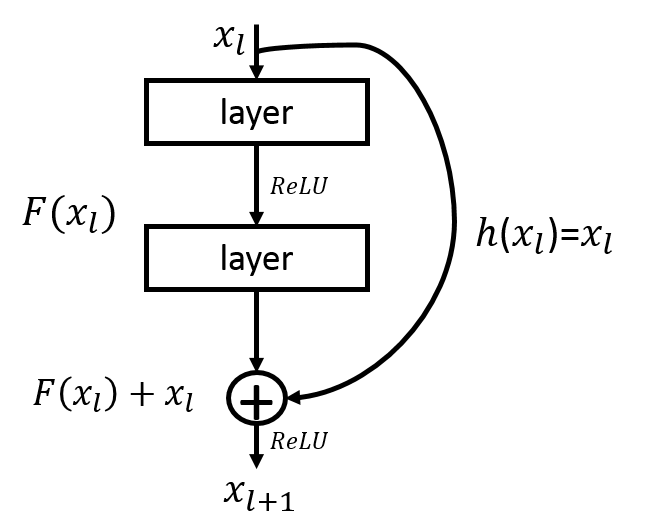}\\
\caption{Residual learning: a building block.}
\label {fig:block4}
\end{figure}

\citet{he2016deep} proposed a deep residual learning framework: let the layers try to learn a residual mapping instead of the directly desired underlying mapping of a few stacked layers. Figure \ref{fig:block4} shows a residual building block. Let the desired underlying mapping be $ H(x_l) $, let the stacked nonlinear layers fit mapping of $F(x_l)=H(x_l)-x_l $. This is residual. The formulation $ F(x_l)=H(x_l)-x_l $ can be written to  $ H(x_l)=F(x_l)+ x_l $ ,  $ F(x_l)+ x_l $ can be realized by feed forward neural networks with \lq\lq short connection\rq\rq~ (Figure \ref{fig:block4}), which skips one or more layers and perform identity mapping. At last, their outputs are added to the outputs of the stacked layers. A residual unit can be expressed as follows:
\begin{equation}\label{eq1}
  x_{l+1}=f(h(x_l)+F(x_l,W_l)).
\end{equation}
Where $ x_l $ and $ x_{l+1} $ are input and output of the $ l$-th  unit, and $ F $ is a residual function. For example, Figure \ref{fig:block4} has two layers, $ F=W_{2}\sigma(W_{1}x) $ in which $ \sigma $ denotes ReLU and the biases are omitted for simplifying notations. $ h(x_l)=x_l $ and $ f $ is a ReLU function.

\begin{figure}
\centering
\includegraphics[scale=0.5]{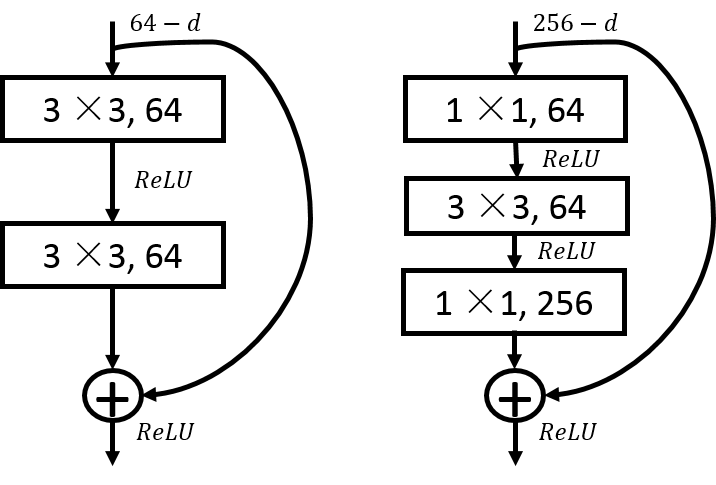}\\
\caption{A deeper residual function F. Left: a building block as in Figure \ref{fig:block4} for ResNet-34. Right: a \lq\lq bottleneck\rq\rq  building block for ResNet-50/101/152/200. Reproduced from Figure 5 in \citet{he2016deep} }\label{fig:block2}
\end{figure}

There are two kinds of residual building blocks in \citet{he2016deep} as shown in Figure \ref{fig:block2}. The basic residual unit (Figure \ref{fig:block2}, left) contains two layers, $ 3\times 3,~ 3\times 3 $ convolutions. In order to decrease the training time and network parameters, a modified residual unit is presented as Figure \ref{fig:block2} (right) shows, which is called a \lq\lq bottleneck\rq\rq~ building block. The \lq\lq bottleneck\rq\rq~ building block uses 3 layers instead of 2 layers and they are $ 1 \times 1,~ 3\times 3,~ 1 \times 1 $ convolutions, where the $ 1 \times 1 $ convolutional layers can reduce and increase dimensions.

In \citet{he2016identity}, both  $ h(x_l) $  and  $ f $ are $ identity~ mapping $, where signal could be directly propagated from one unit to other units, in both forward and backward passes. The residual unit can be redefined as:
\begin{equation}\label{eq3}
  x_{l+1}=x_l+F(x_l,W_l).
\end{equation}

And \citet{he2016identity} also adopted \lq\lq  pre-activation\rq\rq, where \lq\lq BN-ReLU-Conv\rq\rq~ replaced the traditional \lq\lq Conv-BN-ReLU\rq\rq. This is called ResNet V2, which is much easier to train and has better performance than ResNet V1 \citep{he2016deep}. Now, ResNets have many versions, like ResNet-50/101/152/200, deeper layers up to 1001 layers.

\section{Approach} \label{sec:approach}

In the previous section we introduce the theory of ResNets. In this section, we describe our framework including data preprocessing, data augmentation, scale jittering, network architecture, and implementation details.

\subsection{Preprocessing}\label{subsec:preprocessing}

\begin{figure*}
  \centering
  \includegraphics[scale=0.5]{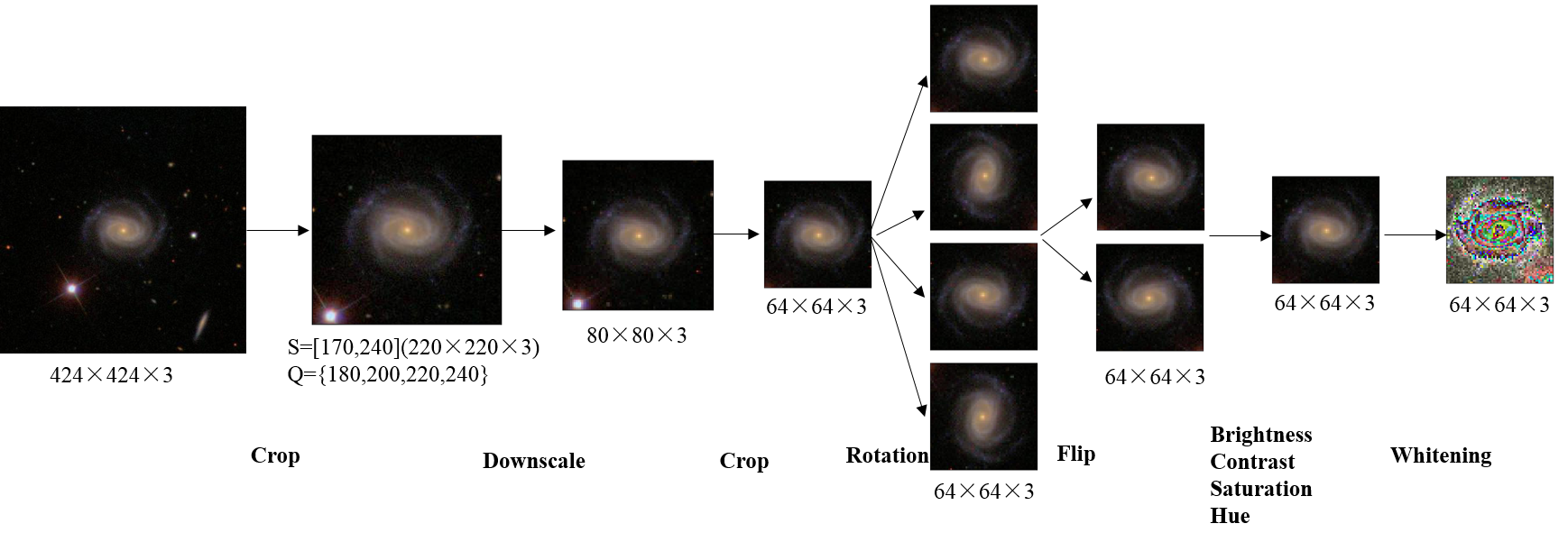}\\
  \caption{Preprocessing procedure. The origan image firstly is center cropped to a range scale $ S=[170,240]$ in training set ($ Q=\{ 180,200,220,240\} $in testing set), for example, the spiral galaxy (GalaxyID:237308) is cropped to $ 220 \times 220 \times 3 $ pixels, then resized to $ 80 \times 80 \times 3 $ pixels, randomly cropped to $ 64 \times 64 \times 3 $ pixels, randomly rotated $ 0^{\circ}, 90^{\circ},180^{\circ},270^{\circ} $, and randomly horizontally flipped. After optical distorting and image whitening, it ($ 64 \times 64 \times 3 $ pixels) becomes the input of networks. }\label{fig:preprocess}
\end{figure*}

From the dataset, images are composed of large fields of view with the galaxy of interest in the center. So it is necessary to crop the image at first step. In practice we crop from the center of image to a range scale $ S=[170,240]$ in training set for every image (as explained later). It allows all the main information to be contained in the center of image, also eliminates many noises like other secondary objects and reduces the dimension of images almost a quarter for faster training. A complete preprocessing procedure is illustrated in Figure \ref{fig:preprocess}.

Then, the image is resized to $ 80 \times 80 \times 3 $ pixels, which is just dimension reduction and easy to compute under limited computing source. Next, a random cropping is carried out, which increases the size of training set by a factor of 256. The size of image drops to $ 64 \times 64 \times 3 $ pixels. Next the image is randomly rotated  with $ 0^{\circ}, 90^{\circ},180^{\circ},270^{\circ} $ because of rotation invariant of galaxy images and randomly horizontally flipped. Brightness, contrast, saturation and hue adjustment are applied to the image and the last step is image whitening. Above is the whole preprocessing pipeline in training. After those steps, images($ 64 \times 64 \times 3 $ pixels) will be used as input of networks when training.

At testing time, preprocessing procedure does not include random cropping, rotation, horizontal flipping and optical distortion. After center cropping to a fixed value $ Q=\{ 180,200,220,240\} $(as explained later), the image is resized to $ 80 \times 80 \times 3 $ pixels and then performs center cropping again, the size of the image is $ 64 \times 64 \times 3 $ pixels. And  the last step is still image whitening, images will be used as input of networks when testing.

\subsection{Data augmentation}\label{subsec:augmentation}

In order to avoid overfitting, data augmentation is one of the common and effective ways to reduce overfitting. Because of our limited training data, data augmentation can enlarge the number of training images. We use five different forms of data augmentation.

Scale jittering is the first form of data augmentation. In training time, we crop the images to a range scale $ S=[170,240]$ , which is called multi-scale training images because of the $ S $ random value. Since different images can be cropped to different sizes and even the same images also can be cropped to different sizes at different iterations, it is beneficial to take this into account during training. This  can be seen as training set augmentation by scale jittering.

Random cropping is carried out from $ 80 \times 80 \times 3 $ pixels to $ 64 \times 64 \times 3 $ pixels, which increases the size of training set by a factor of 256. Rotating training images with $ 0^{\circ}, 90^{\circ},180^{\circ},270^{\circ} $ can enlarge the size of training set by a factor of 4. A horizontal flipping is a doubling of training images.

The first four forms of data augmentation are affine transformations that means the very little computation and they are completed on the CPU before training on the GPUs. Brightness, contrast, saturation and hue adjustment are the same as \citet{krizhevsky2012imagenet}, which are optical distorting for data augmentation.

\subsection{Scale jittering}\label{subsec:jitter}

Scale jittering is derived from \citet{simonyan2014very}, in which images of the input are cropped from multi-scale training images and fixed multi-scale testing images.

\textbf{Training scale jittering}. Let set $ S $ be multi-scale training (we also refer to $ S $ as the training scale), where each training image is individually rescaled by randomly sampling $ S $ from a certain range $ [S_{min}, S_{max}] $ ( we use $ S_{min}=170~ and ~ S_{max}=240 ) $. By this means different images can be cropped to different sizes and even the same images also can be cropped to different sizes at different iterations, that greatly enlarge the number of training set and effectively avoid overfitting. This can be seen as training set augmentation by scale jittering.

\textbf{Testing scale jittering}. Let set $ Q $ be fixed multi-scale testing (we also refer to $ Q $ as the testing scale). In practice, we use $ Q=\{ 180,200,220,240\} $ when testing, which makes our models achieve better performance.

\subsection{Network architecture}\label{subsec:architecture}

\begin{figure}
\centering
\includegraphics[scale=0.5]{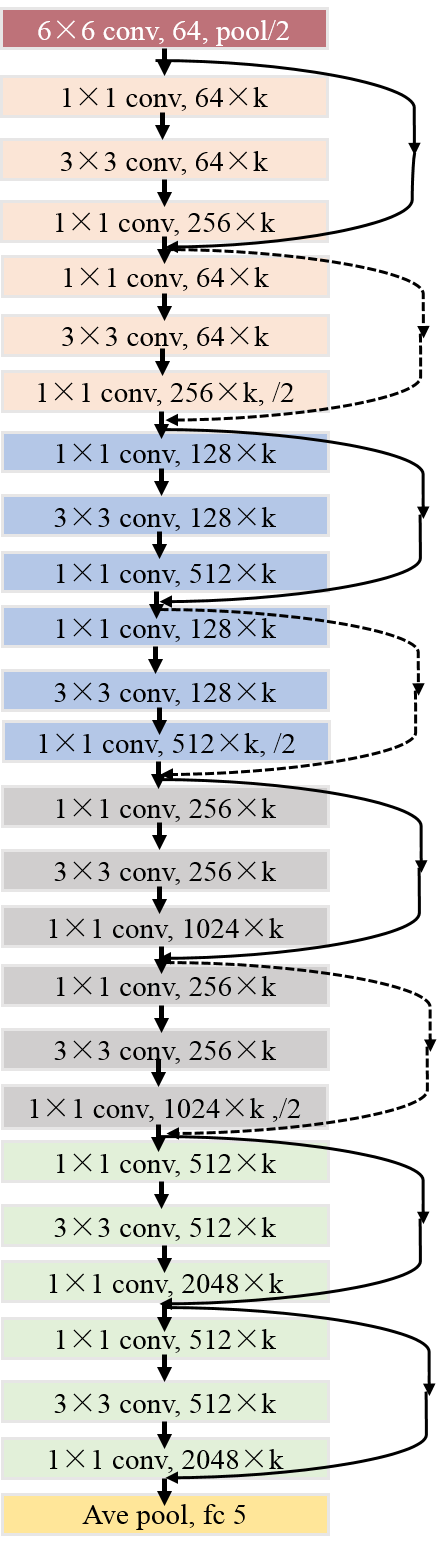}\\
\caption{Our network architecture for Galaxy in this study. where $ k $ is the widening factor. The dashed shortcuts decrease dimensions. Table \ref{tab:model} shows more details.}
\label{fig:galanet}
\end{figure}

Our model is a variant of ResNets V2 \citep{he2016identity}. As Section \ref{subsec:resnet} describes, deep residual networks (ResNets) always seek for deeper and deeper. So the ResNets look like very thin and height. Recent research work shows that such deep residual networks come cross the risk of diminishing feature reuse, which train very slowly and need too much time \citep{zagoruyko2016wide}. We propose a network specially designed for galaxy by trying to decrease the depth and widen residual networks. Our overall architecture of network is depicted in Figure \ref{fig:galanet} and Table \ref{tab:model}.

\begin{figure}
\centering
\includegraphics[scale=0.5]{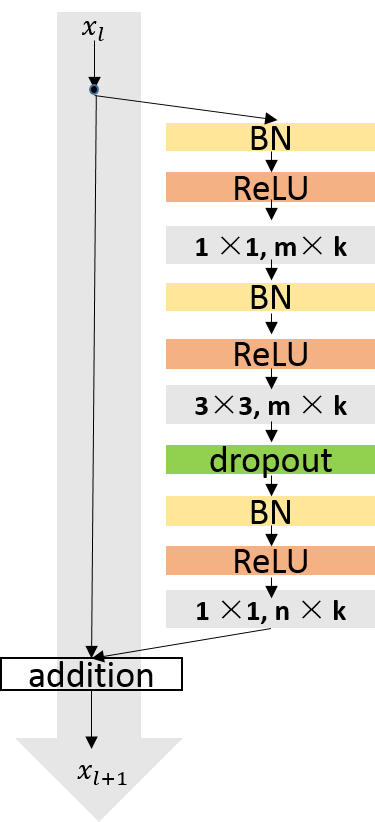}\\
\caption{Full pre-activation residual unit in our study. $ m,~n $ denotes the number of channel, $ k $ is the widening factor. We use $ 1 \times 1,~ 3\times 3,~ 1 \times 1 $ convolutions and the standard \lq\lq BN-ReLU-Conv\rq\rq. }\label {fig:bottleneck}
\end{figure}

We adopt full pre-activation residual units as Figure \ref{fig:bottleneck} shows. And a \lq\lq bottleneck\rq\rq building block( Figure \ref{fig:block2}, right) presented in \citet{he2016deep} is used, namely, a combination of $ 1 \times 1,~ 3\times 3,~ 1 \times 1 $ convolutions, for example, $ 1 \times 1, m \times k ~convolution,~ 3\times 3,m \times k~convolution,~ 1 \times 1,n \times k~convolution $, where $ m,~n $ denotes the number of channel, $ k $ is the widening factor. The full pre-activation includes standard \lq\lq BN-ReLU-Conv\rq\rq. In addition to these, we add a dropout after $ 3\times 3 ~convolution $ whereas ResNet V2 \citep{he2016identity} did not use dropout to prevent coadaptation and overfitting. The residual unit is defined as:

\begin{equation}\label{eq:resnetunit}
  x_{l+1}=x_l+W_{3}\sigma(W_{2}\sigma(W_{1}\sigma(x_l))).
\end{equation}
Here, $ x_l $ and $ x_{l+1} $ are input and output of the $ l$-th unit,  $ \sigma $ denotes BN and ReLU, $ W_{1},~W_{2},~W_{2}$ represent 3 convolutional kernels, dropout is placed after the $ W_{2} $ operation  and the biases are omitted for simplifying notations.

Then looking at our network architecture (Figure \ref{fig:galanet} and Table \ref{tab:model}), the size of input of network is $ 64 \times 64 \times 3 $ pixels. Firstly, 64 kernels of size of $ 6 \times 6 \times 3$ with a stride of 1 are performed, which is derived from \citet{dieleman2015rotation} and proven to be optimal. After the first convolutional layer, a max pooling of size of $ 2\times 2$ with a stride of 2 is connected. The size of output of image becomes $ 32 \times 32 \times 64 $.

The output of max pooling is fed to 4 convolutional groups: conv2, conv3, conv4 and conv5, respectively. Each group has 2 residual blocks. For example, in convolutional group 2, there are 2 residual blocks: $ 1 \times 1, 64 \times 2 $ (128 channels) convolution, $3\times 3,64 \times 2$ (128 channels) convolution, $1 \times 1,256\times 2 $ (512 channels) convolution; $ 1 \times 1, 64 \times 2 $ (128 channels) convolution, $3\times 3,64 \times 2$ (128 channels) convolution, $1 \times 1,256\times 2 $ (512 channels) convolution with a stride of 2, which performs downsampling. Group3, group4 and group 5 are the same, except for the last layer of group 5 does not perform downsampling. Downsampling is performed by the last layers in groups conv2, conv3 and conv4 with a stride of 2.

The dashed shortcuts of Figure \ref{fig:galanet} decrease dimensions. The contributions of $1\times 1$ convolutional layers are reducing dimensions at first and then increasing dimensions, to reduce the parameters of model and speed up training. The last layer is global average-pooling layer with $ 4 \times 4 $ kernel and the size of output of average pooling is $ 1 \times 1 \times 4096 $. At last is a 5-way fully connected layer with $ softmax $.

Where $ k $ is the widening factor, $ N $ denotes the number of blocks in group. After hundreds of trying, we  finally use $k=2$, $N=2$ in practice. So our network is 26 layers totally including 26.3M parameters. The 26-layers network achieves the best performance on accuracy and other metrics.

From our network architecture, some tips are concluded: the first convolutional layer adopts a relatively large convolution filter of $ 6 \times 6 $; the convolutional layers mostly have $ 1 \times 1$ and $ 3\times 3 $ convolutions. The advantages of $ 1 \times 1$ convolutions have been described. The advantages of small  $ 3\times 3 $ filter have been demonstrated in  \citep{simonyan2014very}, which can decrease the number of parameters of model and achieve a better performance. The feature maps in each group are the same except the last layer of each convolutional group(conv2, conv3 and conv4). The feature map size is halved, the number of filters is doubled.

\begin{table}
\centering
\caption{Architecture of our model for Galaxy in this study. Residual units are shown in brackets. where $ k $ is the widening factor, $ N $ denotes the number of blocks in group (We use $k=2$, $N=2$, which means our network is 26 layers totally). Downsampling is performed by the last layers in groups conv2, conv3 and conv4 with a stride of 2.}
\label{tab:model}
\begin{tabular}{c|c|c}
  \hline\hline
  Layer name &Output size & Depth \\\hline
  Conv 1 & $64\times64$ &$ 6\times6, 64$  \\
  Max-pooling & $32\times32 $ & $ 2\times 2 $, stride~2 \\
  Conv 2& $ 16\times16 $ & $\begin{bmatrix}1\times 1,64 \times k\\3 \times 3, 64 \times k\\1 \times 1,256\times k\end{bmatrix}\times N $ \\
  Conv 3 &$ 8\times8 $  & $\begin{bmatrix}1\times 1,128 \times k\\3 \times 3, 128 \times k\\1 \times 1,512\times k\end{bmatrix}\times N $  \\
  Conv 4 & $ 4\times4$  & $\begin{bmatrix}1\times 1,256 \times k\\3 \times 3, 256 \times k\\1 \times 1,1024\times k\end{bmatrix}\times N $  \\
  Conv 5 & $4\times4$ & $\begin{bmatrix}1\times 1,512 \times k\\3 \times 3, 512\times k\\1 \times 1,2048\times k\end{bmatrix}\times N $  \\
  Avg-pooling & $1\times1$ & $ 4\times4,5-d,softmax $ \\
  \hline
\end{tabular}
\end{table}

\subsection{Implementation Details}\label{subsec:implementation}

We use mini-batch gradient descent with a batch size of 128 and Nesterov momentum of 0.9. The initial learning rate is set to 0.1, then decreased by a factor of 10 at 30k and 60k iterations, and we stop training after 72k iterations. The weight decay is 0.0001, dropout probability value is 0.8 and the weights are initialized as in \citet{he2015delving}. We adopt BN before activation and convolution, following \citet{he2016identity}.

Our implementation is based on Python, Pandas, scikit-learn \citep{Pedregosa2012Scikit}, scikit-image \citep{Van2014scikit}, TensorFlow \citep{abadi2016tensorflow}. It takes  about 31.5 hours to train a single network  with a  NVIDIA Tesla K80 GPU. Our code is available at \url{https://github.com/Adaydl/GalaxyClassification}.

\section{Results and Discussion} \label{sec:results}

In this section, we describe 7 kinds of classification performance metrics: accuracy, precision, recall, F1, confusion matrix, ROC and AUC. Then we show the results of our model and compare systematically the performance of our model with other popular CNNs models, such as Dieleman, AlexNet, VGG, Inception and ResNets. In the end we visualize the filters and feature maps.

\subsection{Classification Performance Metrics}\label{subsec:metrics}

To assess the performance of our classification models, we present 7 kinds of classification performance metrics: accuracy, precision, recall, F1, confusion matrix, ROC and AUC. They are defined as follow:

\textbf{Accuracy}: $\hat{y_i}$ is the predicted value of the $i$-th sample and $ y_i$ is the corresponding true value, then the fraction of correct predictions over $ n_{samples}$ is defined as

\begin{equation}\label{acc}
  Accuracy(y_i,\hat{y_i})=\frac{1}{n_{samples}}\sum_{i=0}^{n_{samples}-1}1(\hat{y_i}=y_i).
\end{equation}

\textbf{Precision}, \textbf{Recall} \&\textbf{ F1} \citep{Ceri2013}: Given the number of true positive (TP), false positive (FP), true negative (TN) and false negative (FN), we define:

\begin{equation}\label{p}
  P=\frac{TP}{TP+FP}.
\end{equation}

\begin{equation}\label{r}
  R=\frac{TP}{TP+FN}.
\end{equation}

\begin{equation}\label{F1}
  F1=\frac{2PR}{P+R}.
\end{equation}

\textbf{Confusion Matrix(CM)}: An entry $ CM_{ij} ( i,j=1,2, \cdots, n_{samples}) $ is defined as the number of the true class $ i $ ,but predicted to class $ j $.

\textbf{ROC} \& \textbf{AUC}: A receiver operating characteristic (ROC) curve plots the true positive rate against the false positive rate for every possible classification threshold. AUC is the area under the receiver operating characteristic (ROC) curve. The closer the AUC is to 1, the better the classification performance.

\subsection{Classification Results and Discussion}\label{subsec:results}

In this section, we summary the results of our models on 7 kinds of classification performance metrics and compare the results of our model with other popular CNNs.

Table \ref{tab:prf} shows that precision, recall and F1 of our model for each class on testing set. 0, 1, 2, 3 and 4 represent completely round, in-between, cigar-shaped, edge-on and spiral, respectively. The average precision, recall and F1 of the 5 classes galaxies of our model are 0.9512, 0.9521 and 0.9515. The completely round achieves the best precision of 0.9611. The spiral achieves the best recall of 0.9782 and F1 value of 0.9677. On the whole, the results of the completely round, the in-between, the edge-on and the spiral are extremely excellent, except the cigar-shaped. It happens due to the small number of the cigar-shaped images for training.

The confusion matrix of our model for each class on testing set is shown in Table \ref{tab:cm}. Column represents true label and row represents prediction label. 815 completely round, 762 in-between, 34 cigar-shaped, 368 edge-on and 763 spiral are classified correctly. So the accuracy of the 5 galaxy types are: completely round, 96.6785\%; in-between, 94.4238\%; cigar-shaped, 58.6207\%; edge-on, 94.3590\% and spiral, 97.6953\% respectively. 29 completely round are incorrectly classified as in-between. It is common sense that completely round and in-between are similar itself and easily misclassified. Note that 4 completely round are misclassified as spiral, perhaps due to faint images photographed from far away distance.  It observes that 12 cigar-shaped are misclassified as edge-on and 18 edge-on are misclassified as cigar-shaped, where the number of misclassifications is greater than others.  We suppose that it happens due to the similarity of cigar-shaped and edge-on, which is so surprising.

\begin{table}
\centering
\caption{Precision, Recall and F1 of our model for each class on testing set.}
\label{tab:prf}
\begin{tabular}{cccc}
  \hline\hline
  Class & Precision & Recall & F1  \\\hline
  0 & \textbf{0.9611} & 0.9634 & 0.9622 \\
  1 & 0.9561 & 0.9431 & 0.9495  \\
  2& 0.7234 &0.5862  & 0.6476 \\
  3 & 0.9412 & 0.9485 & 0.9448  \\
  4 & 0.9573&  \textbf{0.9782}& \textbf{0.9677}  \\
  Average & 0.9512 & 0.9521 &0.9515 \\
  \hline
\end{tabular}
\end{table}

\begin{table}
\centering
\caption{Confusion matrix of our model for each class on testing set. Column represents true label and row represents prediction label. }
\label{tab:cm}
\begin{tabular}{c|ccccc}
  \hline
   & 0 & 1 & 2 & 3 & 4 \\
  \hline
  0 & \textbf{815} & 21 & 0 & 0 & 10\\
  1 & 29 & \textbf{762} & 0 & 0 & 17\\
  2 & 0 & 4 & \textbf{34} & 18 & 2 \\
  3 & 0& 3 &12  & \textbf{368} & 5 \\
  4 & 4 & 7 & 1 & 5 & \textbf{763} \\
  \hline
\end{tabular}
\end{table}

\begin{figure}
  \centering
  \includegraphics[scale=0.5]{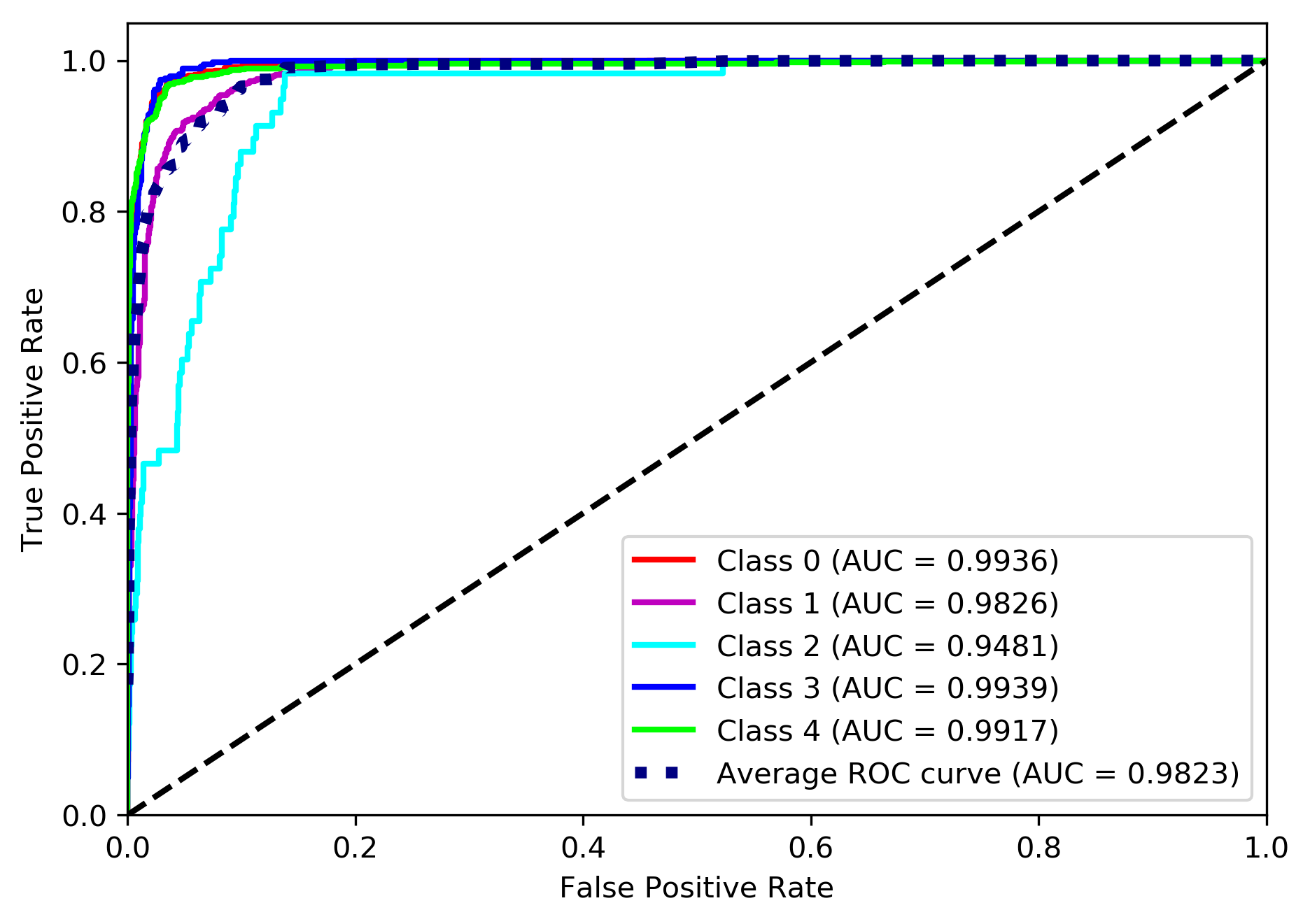}\\
  \caption{ROC curve of our model for 5 classes galaxies on testing set. Each color represents a class.}\label{fig:roc}
\end{figure}

Figure \ref{fig:roc} shows that ROC curve of our model for 5 classes galaxies on testing set. Each color represents a class. The closer the true positive rate (TPR) is to 1 and false positive rate (FPR) is to 0, the better the curve predicts, namely, the closer the curve is to the upper left corner, the better it predicts. From Figure \ref{fig:roc},  ROC curve of each class performs well, the edge-on predicts the best and the cigar-shaped predicts relatively worse, which happens due to the small number of cigar-shaped images. The average AUC of our model is 0.9823, and shows that the overall prediction performance of our model is excellent.

Table \ref{tab:acc} summaries  test accuracy of different methods at multiple test scales. Our results are based on average values of the maximum values of 10-times runs of each test scale. Recent research work shows scale jittering at testing time can obtain a better performance \citep{simonyan2014very}. Our model obtain the best results with 94.6875\% accuracy. Table \ref{tab:acc} shows that Dieleman model \citep{dieleman2015rotation} works well, obtains 93.8800\% accuracy, although it is a only 7-layers CNN. It is easy to understand that it is designed specifically for galaxy images and other networks, such as AlexNet \citep{krizhevsky2012imagenet}, VGG \citep{simonyan2014very}, Inception \citep{szegedy2015going, ioffe2015batch, szegedy2016rethinking, szegedy2017inception} and ResNets \citep{he2016deep, he2016identity}, are designed for ImageNet, but they all have excellent performance because of their good generalization performance. AlexNet is a 8-layers CNN won the first place in the ImageNet LSVRC-2012 in 2012 years, here, it achieves a 91.8230\% accuracy due to its used relatively large filter( $ 11 \times 11 $ convolution). VGG-16 achieves a 93.1336\% accuracy, which uses many small $ 3 \times 3 $ filters. Inception here implemented is Inception V3 including 42 layers with careful designed inception module, and here achieves 94.2014\% accuracy. ResNet-50 here implemented is pre-act-ResNets and  obtains a 94.0972\% accuracy.

\begin{table*}
\centering
\caption{Test accuracy of different methods at multiple testing scales. Our results are based on average values of the maximum values of 10-times runs of each testing scale. The bold entries highlight the best results.}
\label{tab:acc}
\begin{tabular}{c|c|c|c}
  \hline
  \hline
  \multirow{2}{*}{Model}& \multicolumn{2}{c|}{Image side} &\multirow{2}{*}{Accuracy(\%)}  \\
  \cline{2-3}
  &Train(S)  & Test(Q) &  \\
  \hline
  Dieleman\citep{dieleman2015rotation} &[170,240]  & 180,200,220,240 & 93.8800 \\
   AlexNet\citep{krizhevsky2012imagenet}& [170,240] & 180,200,220,240 & 91.8230 \\
   VGG\citep{simonyan2014very}& [170,240] & 180,200,220,240 & 93.1336 \\
   Inception\citep{szegedy2016rethinking}& [170,240] & 180,200,220,240 & 94.2014 \\
   ResNet-50\citep{he2016identity}& [170,240] & 180,200,220,240 & 94.0972 \\
   Ours& [170,240] & 180,200,220,240 & \textbf{94.6875} \\
  \hline
\end{tabular}
\end{table*}

Table \ref{tab:metric} summaries  test accuracy, precision, recall, F1 and AUC of different methods. Our results are based on the maximum values of 10-times runs of each testing scale. We notice the results of accuracy are better than the results indicated in Table \ref{tab:acc}, because they are obtained by picking the maximum values of 10-times runs of each testing scale, instead of the average values of the maximum values of 10-times runs of each testing scale. Our model achieves the best accuracy 95.2083\% at single testing scale. Because accuracy has a fatal flaw in multi-class task that it depends on the number of the majorities, so we also adopt average precision, recall, F1 and AUC to measure classification performance. Our model obtains the best average precision 0.9512, the best average recall 0.9521 and the best average F1 0.9515.  Inception achieves the best average AUC 0.9852. On the whole, our model works excellent and achieves state-of-the-art performance.

\begin{table*}
\centering
\caption{Test accuracy, precision, Recall, F1 and AUC of different methods. Our results are based on the maximum values of 10-times runs of each testing scale. The bold entries highlight the best results within each column.}
\label{tab:metric}
\begin{tabular}{cccccc}
  \hline
  \hline
  Model & Accuracy(\%) & Precision & Recall & F1 & AUC \\\hline
  Dieleman\citep{dieleman2015rotation}&94.6528 & 0.9455 & 0.9465 & 0.9456 & 0.9793   \\
  AlexNet\citep{krizhevsky2012imagenet} & 92.2569 & 0.9207 & 0.9226 & 0.9215 & 0.9809 \\
  VGG\citep{simonyan2014very} & 93.6458 & 0.9348 & 0.9365 & 0.9353 & 0.9846 \\
  Inception\citep{szegedy2016rethinking} & 94.5139 & 0.9447 & 0.9451 & 0.9448 & \textbf{0.9852} \\
  ResNet-50\citep{he2016identity} & 94.6875  & 0.9458 & 0.9469  & 0.9461 & 0.9823  \\
  Ours & \textbf{95.2083}  & \textbf{0.9512} & \textbf{0.9521}  & \textbf{0.9515} & 0.9823 \\
  \hline
\end{tabular}
\end{table*}

\subsection{Filters and Feature Maps Visualization}\label{subsec:visualization}

Neural networks are always known as \lq\lq black boxes\rq\rq. We want to visualize what the CNN learn by visualizing filters weights and feature maps and then give a qualitative empirical analysis\citep{zeiler2014visualizing,yosinski2015understanding}. In order to understand easily, we visualize a simple CNN, 7 layers totally, including 4 convolutional layers ($6\times 6 $, 32 filters, $ 5\times 5$, 64 filters, $ 3 \times 3 $, 128 filters, and $ 3\times 3 $, 128 filters, respectively) and 3 fully connected layers.

\begin{figure}
  \centering
  \includegraphics[scale=0.8]{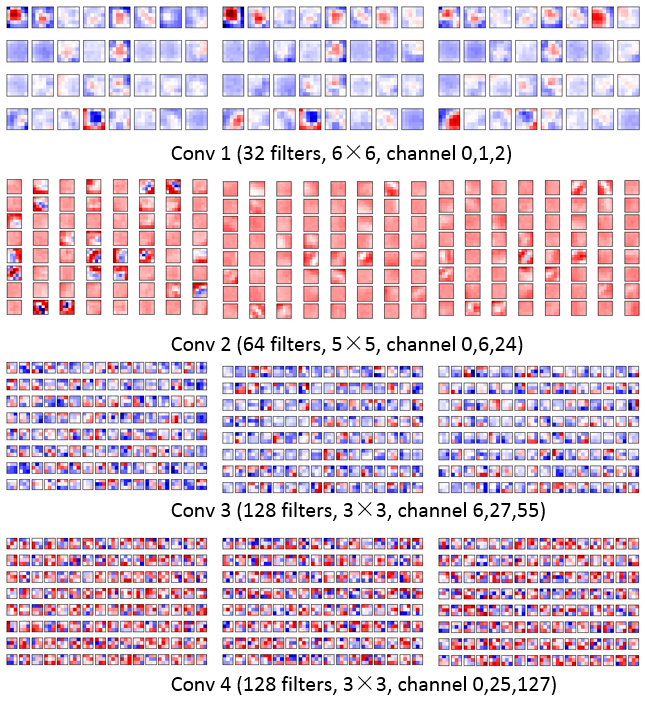}\\
  \caption{Filter weights learned on every convolutional layer. From top to bottom, they are filter weights of 4 convolutional layers. From left to right, they are filter weights visualization of different channels on certain convolutional layer. Brackets show the number of filters, the size of filters and channels visualized. }\label{fig:filter}
\end{figure}

Figure \ref{fig:filter} shows that filter weights learned on every convolutional layer. The first layer filters detect the different galaxy edges, corners, etc. from original pixel, then use the edge to detect simple shapes in second layer filters, such as the bar, the elliptical and so on, and then use these shapes to detect more advanced features in high level layer filters. More invariant representations are learned with the increase of layers.  And from Figure \ref{fig:filter}, different filters also learn different color information, mainly red and blue that might correspond to the color of galaxy itself, such as red elliptical galaxy and blue spiral galaxy.

\begin{figure}
  \centering
  \includegraphics[scale=0.60]{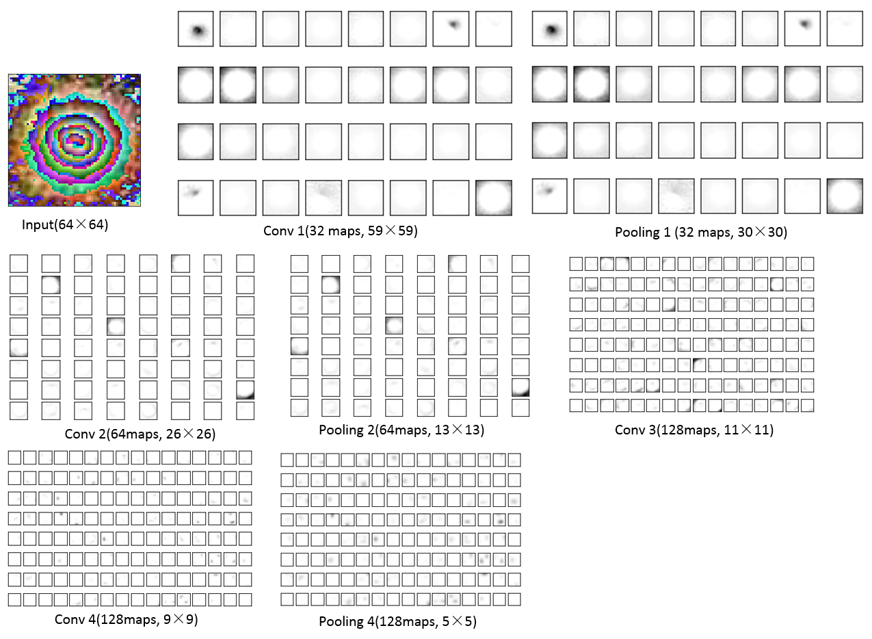}\\
  \caption{Activations of of each layer on a smooth galaxy(GalaxyID: 909652). From top to bottom, left to right, they are: input image after whitening; Activations on the Conv 1, Pooling 1, Conv 2, Pooling 2, Conv 3, Conv 4 and Pooling 4. Brackets show the number of feature maps and the size of feature maps.}\label{fig:smooth}
\end{figure}

\begin{figure}
  \centering
  \includegraphics[scale=0.60]{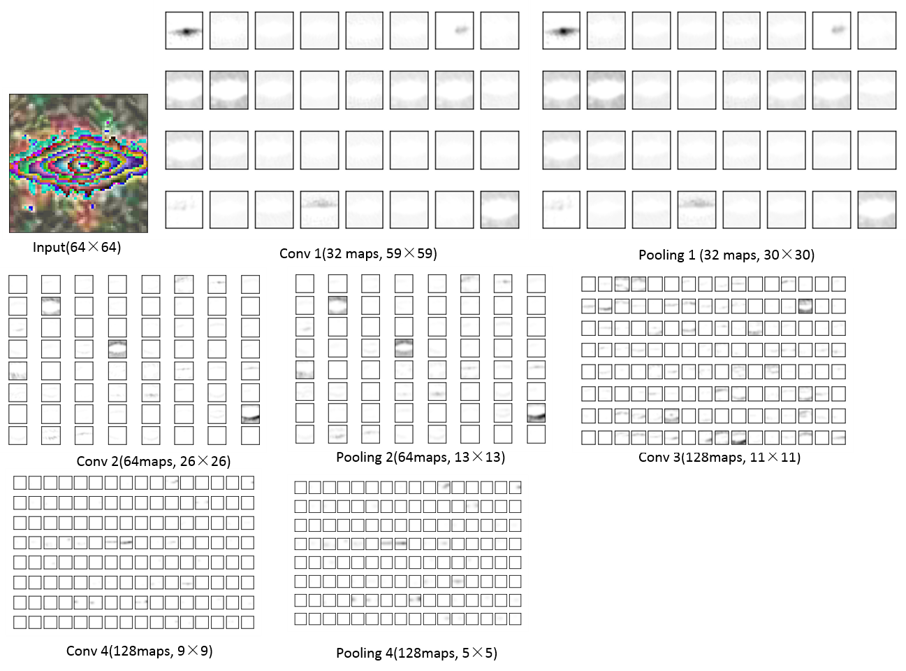}\\
  \caption{Similar to Figure \ref{fig:smooth} but for an edge-on galaxy(GalaxyID: 416412).}\label{fig:edge}
\end{figure}

\begin{figure}
  \centering
  \includegraphics[scale=0.58]{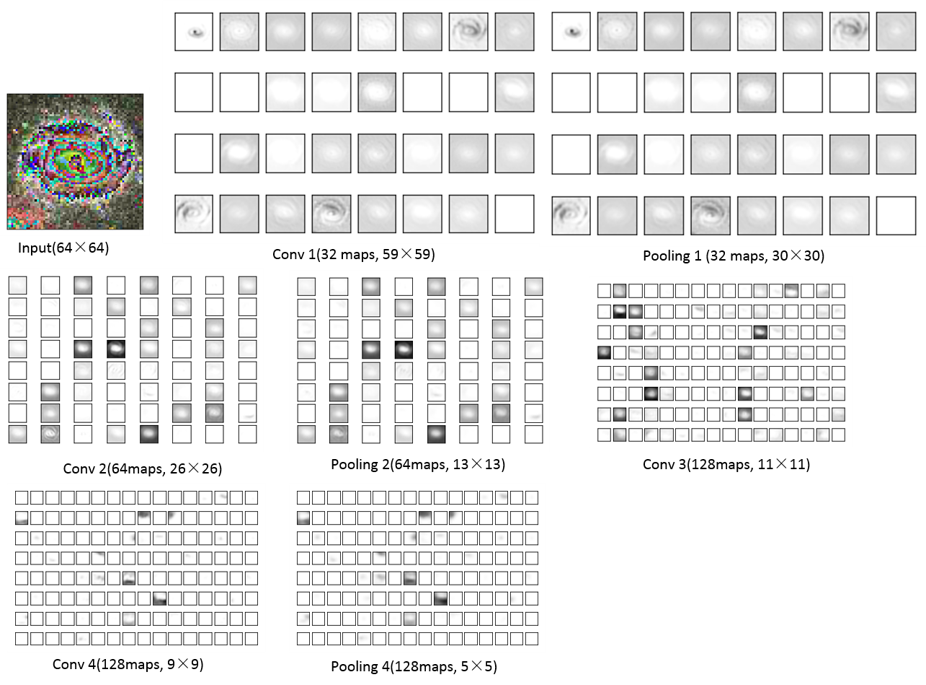}\\
  \caption{Similar to Figure \ref{fig:smooth} but for a spiral galaxy(GalaxyID: 237308).}\label{fig:spiral}
\end{figure}

Figure \ref{fig:smooth} shows that activations of of each layer on a smooth galaxy (GalaxyID: 909652). In first layer, some feature maps recognize the intermediate core of galaxy, and some recognize the background part. In high layers, feature maps recognize the abstract blobs with the combination of high-level features, e.g., in the fourth convolutional layer.
It is seen that after pooling layers, the differentiability of each feature map is stronger, which is exactly what the classification model expects. These interesting phenomenons also can be found in  Figure \ref{fig:edge} and Figure \ref{fig:spiral}.

\section{Conclusions } \label{sec:conclusions}

In this paper, we propose a variant of residual networks (ResNets) for Galaxy morphology classification. We classify 28790 galaxies into 5 classes, namely, completely round smooth, in-between smooth (between completely round and cigar-shaped), cigar-shaped smooth, edge-on and spiral using Galaxy Zoo 2 dataset. In data preprocessing, a complete preprocessing pipeline is presented and five forms of data augmentation are adopted to avoid overfitting, especially scale jittering that extremely enlarges the number of training images.

The advantage of our network is combining Dieleman model with residual networks (ResNets), in which we try to decrease the depth and widen residual network. We use a \lq\lq bottleneck\rq\rq  residual unit with full pre-activation \lq\lq BN-ReLU-Conv\rq\rq. In order to ovid overfitting, we use dropout after $ 3\times 3 ~convolution $. Our network has 26 layers with 26.3M parameters. We make a systematic comparation between our model and other popular convolutional networks (CNNs) in deep learning, such as Dieleman, AlexNet, VGG, Inception and ResNets. Our model achieves the best classification performance, the overall accuracy on testing set is 95.2083\% and the accuracy of the 5 galaxy types are: completely round, 96.6785\%; in-between, 94.4238\%; cigar-shaped, 58.6207\%; edge-on, 94.3590\% and spiral, 97.6953\% respectively. The average precision, recall, F1 and AUC of our model are 0.9512, 0.9521, 0.9515 and 0.9823. From the confusion matrix, we find that 12 cigar-shaped are misclassified as edge-on and 18 edge-on are misclassified as cigar-shaped, where the number of misclassifications is greater than others. We suppose that it happens due to the similarity of cigar-shaped and edge-on, which is so surprising. Dieleman model also works well and the average accuracy is 94.6528\% because it is specially designed for galaxy images. Although AlexNet, VGG, Inception and ResNets are designed for ImageNet, they all achieve excellent performance because of their good generalization, whose accuracies are: 92.2569\%, 93.6458\%, 94.5139\% and 94.6875\%.

By visualizing filters weights and feature maps, we try to understand what the CNN model learn. For instance, the first layer filters detect the different galaxy edges, corners, etc. from original pixel, then use the edge to detect simple shapes in second layer filters, such as the bar and the elliptical, and next use these shapes to detect more advanced features in high level layer filters. We also find that different filters also learn different color information, mainly red and blue, which might correspond to the color of galaxy itself, such as red elliptical galaxy and blue spiral galaxy. About activations of of each layer on a galaxy image, some feature maps recognize the intermediate core, some recognize the background part in the first layer, feature maps recognize the abstract blobs with the combination of high-level features in higher layers. It also is found that after pooling layers, the differentiability of each feature map is stronger, which is exactly what the classification model expects.

In future large-scale surveys, such as the Dark Energy Survey (DES) and the Large Synoptic Survey Telescope (LSST), will obtain billions of galaxy images and our algorithms can be applied to automatically classify galaxies and achieve sate-of-the-art performance.

In future work, we focus on much more fine-grained galaxy morphology classification. We plan to train our model on bigger and higher quality galaxy dataset. In the end, more advanced algorithms in deep learning will merge with galaxy morphology classification.

\section*{Acknowledgements}

We would like to thank the galaxy challenge, Galaxy Zoo, SDSS and Kaggle platform for sharing data. We acknowledge the financial support from the National Earth System Science Data Sharing Infrastructure (\url{http://spacescience.geodata.cn}). We are supported by  CAS e-Science Funds (Grand XXH13503-04).




\bibliographystyle{mnras}
\bibliography{cit} 





\bsp	
\label{lastpage}
\end{document}